\documentclass[final,11pt,times]{elsarticle}
\usepackage{graphicx}
\usepackage{eurosym}
\usepackage{amsmath}
\usepackage{amssymb}
\usepackage[colorlinks,urlcolor=blue]{hyperref}
\usepackage{geometry}

\biboptions{sort&compress}
\journal{CSF}

\begin{document}

\begin{frontmatter}
\title{\textbf{Three-dimensional solitons supported by the spin-orbit coupling and Rydberg-Rydberg interactions in $\mathcal{PT}$-symmetric potentials}}
 \author{\textbf{Yuan Zhao}$^{1,2\&}$}
\author{\textbf{Qihong Huang}$^{1,3\&}$}
\author{\textbf{Tixian Gong}$^{1,3}$}
\author{\textbf{Siliu Xu}$^{1,2*}$ }
\author{\textbf{Zeping Li}$^{1,3*}$ }
\author{\textbf{Boris A. Malomed}$^{4,5}$ }
 \cortext[cor]{Corresponding authors: xusiliu1968@163.com (Siliu Xu) and 18271438135@163.com (Zeping Li)}

\address{$^{1}$Key Laboratory of Optoelectronic Sensing and Intelligent Control, Hubei University of Science and Technology, Xianning, 437100, China}
\address{$^{2}$School of Biomedical Engineering and Imaging, Xianning Medical College, Hubei University of Science and Technology, Xianning 437100, China}
\address{$^{3}$School of Electronic and Information Engineering, Hubei University of Science and Technology, Xianning 437100, China}
\address{$^{4}$Department of Physical Electronics, School of Electrical Engineering, Faculty of Engineering, Tel Aviv University, P.O.B. 39040, Ramat Aviv, Tel Aviv, Israel}
\address{$^{5}$ Instituto de Alta Investigaci\'on, Universidad de Tarapac\'a, Casilla 7D, Arica, Chile}
\address{$^{\&}$ These authors contributed equally in this work}
\begin{abstract}
Excited states (ESs) of two- and three-dimensional (2D and 3D) solitons of the semivortex (SV) and mixed-mode (MM) types, supported by the interplay of the spin-orbit coupling (SOC) and local nonlinearity in binary Bose-Einstein condensates, are unstable, on the contrary to the stability of the SV and MM solitons in their fundamental states. We propose a stabilization strategy for these states in 3D, combining SOC and long-range Rydberg-Rydberg interactions (RRI), in the presence of a spatially-periodic potential, that may include a parity-time ($\mathcal{PT}$)-symmetric component. ESs of the SV solitons, which carry integer vorticities $S$ and $S+1$ in their two components, exhibit robustness up to $S= 4$. ESs of MM solitons feature an interwoven necklace-like structure, with the components carrying opposite fractional values of the orbital angular momentum. Regions of the effective stability of the 3D solitons of the SV and MM types (both fundamental ones and ESs), are identified as functions of the imaginary component of the $\mathcal{PT}$-symmetric potential and strengths of the SOC and RRI terms.

\end{abstract}


\end{frontmatter}

\section{Introduction}

The formation of multidimensional solitons holds fundamental significance
across various domains of physics, including nonlinear optics \cite%
{optics-Kartashov-2019,optics-Mihalache-2024,optics1creating2001ruostekoski,optics2evidence2009shomroni,optics3stable2012takuto}%
, Bose-Einstein condensates (BECs) \cite%
{BECs2015-Bagnato,BECs12008PG,BECs22009Fetter,BECs32004Eiermann}, superconductors \cite%
{superconductordual2002babaev}, semiconductors \cite%
{semiconductor2009Topological}, ferromagnetic media \cite%
{Cooper2008Propagating}, and general field theory \cite%
{fieldtheory1-1999,fieldtheory2-2008}.
Unlike one-dimensional (1D) solitons,
which emerge, normally, as stable modes, looking for stable 2D and,
especially, 3D solitons is a challenging problem \cite{book}. Indeed, the
prerequisite for the formation of solitons is the presence of an attractive
(or self-focusing) nonlinearity. However, the ubiquitous cubic
self-attraction gives rise to the destabilizing critical and supercritical
wave collapse in the 2D and 3D geometry, respectively \cite%
{wavecollapse1-1998Wave,wavecollapse2-2011kuznetsov}. Solitons with embedded
vorticity are subject to a still stronger splitting azimuthal modulational
instability \cite{intrinsicvorticity2000kivshar}.

Various stabilization scenarios for fundamental and vortex multidimensional
solitons have been elaborated \cite{book}. These include the use of
saturable or competing nonlinearities \cite%
{nonlinearity1-2000Three,nonlinearity2-2002Stable}, nonlocal nonlinearity
\cite{nonlocal1-2011Rydberg,nonlocal2-2008Anisotropic}, and special
nonlinear interactions \cite{special1-2011Yaroslav,special2-2013Robust}.
Additionally, the stabilization of multidimensional solitons may occur in
optical tandem systems \cite{tandem2009Light}, waveguide arrays and optical
lattices embedded in various materials \cite%
{array2019Dong,array1-2004Multidimensional,array2-2005Stable,array3-2018Bessel,array4-2023Bessel2}, and long-range interactions
between Rydberg atoms \cite{nonlocal1-2011Rydberg,Rydberg2019BaiZhengyang}.
More recently, stable 2D and 3D solitons were predicted, respectively, as
ground \cite{SOC1-2013Spin,SOC2-2018Liyongyao,SOC3-2020LiYongyao} or
metastable \cite{SOC2015Stable} states in binary Bose-Einstein condensates
(BECs) carrying spin-orbit coupling (SOC).

SOC BECs, experimentally realized in Ref. \cite{experiment-Lin2011Spin},
have attracted much interest due to their potential for realizing phenomena
induced by the artificial gauge potential \cite{gauge-2011Colloquium}. 
These settings give rise to various matter-wave modes, such as vortices \cite%
{vortices1-2011Spin,vortices2-2023Liyongyao}, monopoles \cite%
{monopole2012Line}, multi-domain patterns \cite{multi-domain2010Spin}, quantum droplets \cite{quantum droplets1,quantum droplets2,quantum droplets3,quantum droplets4,quantum droplets5}, and
solitons \cite{soliton1-2013Matter}-\cite{EPL}. 
Further, Ref. \cite%
{potential1-2014Creation} has predicted various self-trapped vortex-soliton
complexes in the 2D binary-BEC model with the Rashba-type SOC and attractive
intrinsic nonlinearity. Refs. \cite{potential2-2014Bose} and \cite%
{localized-2D} have used systems with spatially confined SOC in BEC as
localized gauge potentials, examining properties of the respective soliton
complexes and spinor dynamics. As demonstrated in Ref. \cite{SOC2015Stable},
a self-attractive binary SOC condensate can sustain metastable 3D solitons
in the free space, in spite of the absence of a ground state, which does not
exist in the system admitting the supercritical collapse. The study
indicates that the SOC-induced change of the 3D condensate's dispersion may
balance the attractive nonlinearity, resulting in the formation of the
solitons which are stable against small perturbations.

A generic approach relies on the use of spatially periodic potentials, which
secure the stabilization of multidimensional solitons \cite%
{periodic1-2009Observation,periodic2-2008Discrete}. Furthermore, in the case
of the local self-repulsive nonlinearity, stable gap solitons exist in
spectral bandgaps created by such periodic potentials \cite%
{BECs32004Eiermann,Thawatchai,gap1-exciton-polariton2013e}. The
stabilization of solitons can be enhanced by fine-tuning the nonlinearities
and magnitude of the external potentials. Notably, the incorporation of
parity-time ($\mathcal{PT}$) symmetry into periodic potentials has been
demonstrated as an effective technique for generating a variety of optical
solitons. This is achieved by adding an imaginary spatially-odd component to
the spatially-even real periodic potential \cite%
{PT2019Dong,parity-time1-2018chao,parity-time-2023zhao,symm-breaking,parity-time2-2016}. Recently, Ref. \cite%
{kartashov-three-dimensional2016} addressed 3D solitons in complex $\mathcal{%
PT}$-symmetric periodic lattices combined with the focusing Kerr
nonlinearity. It was found that such lattices are capable to stabilize both
fundamental and vortex soliton states.

Cold Rydberg atoms offer a powerful platform for the generation of stable
multidimensional solitons, both fundamental and vortical ones \cite%
{vortices1-2011Spin,vortices2-2023Liyongyao}. The long-range Rydberg-Rydberg
interaction (RRI) gives rise to a nonlocal optical nonlinearity via the
effect of the electromagnetically induced transparency, producing a
significant impact even at the single-photon level \cite%
{parity-time1-2018chao}. This line of the studies has led to the prediction
of numerous optical phenomena, including stable storage and retrieval of
light bullets \cite{Rydberg2019BaiZhengyang}, their two-component
counterparts \cite{two-component2022Zhao}, controllable bound states of the
bullets \cite{LBs-stable2022lu}, transient optical responses \cite%
{transient-2020}, and effective photon interactions \cite%
{photon-interacting2016yang}. The extended lifetimes of Rydberg atomic
states, on the order of tens of microseconds, secure the coherence of the
resultant optical nonlinearities. These characteristics position Rydberg
systems as nearly ideal candidates for the development of advanced photonic
devices, such as single-photon switches and transistors \cite%
{switch1-2014Single,switch2-2016Enhancement}, quantum memory units, and
phase gates \cite{memory1-2013Storage,memory2-2016Quantum}.

Despite recent advancements, 3D vortex solitons in binary condensates, which
are supported by the interplay of SOC and RRI in the presence of the
spatially potentials, real or $\mathcal{PT}$-symmetric ones, are not yet
fully understood. The objective of the present work is to systematically
construct 3D stable solitons in the binary SOC BEC, in conjunction with
mean-field interactions and $\mathcal{PT}$-symmetric potentials, through a
comprehensive numerical analysis. We demonstrate the existence of 3D stable
solitons of the semivortex (SV) and mixed-mode (MM) types. In particular,
for the 3D SV solitons with integer vorticities $\left( S,S+1\right) $ in
their two components, such results are reported for $S\leq 4$, i.e.,
essentially, for higher-order solitons (alias excited states (ESs) of the
fundamental solitons, which correspond solely to $S=-1$ and $S=0$). Our
findings reveal that the imaginary component of the $\mathcal{PT}$ symmetry
potential, along with the SOC and RRI strengths significantly affect the
structure and stability domains of the two types (SV and MM) of the
higher-order solitons.

\section{The spin-orbit-coupled condensate of Rydberg atoms}

In this work, we aim to predict robust SV and MM solitons in the 3D SOC BEC
system, under the action of a $\mathcal{PT}$-symmetric potential. To this
end, we consider matter waves in Rydberg atomic gases, where the interplay
between the $\mathcal{PT}$ symmetry and RRI may give rise to novel soliton
phenomenology. With the help of a systematic numerical analysis, we explore
the shape and stability of the solitons.

The mean-field dynamics of the two-component wave function of the binary BEC
is governed by the spinor Gross-Pitaevskii equation (GPE),
\begin{gather}
i\hbar \frac{\partial \psi _{+}}{\partial t}=\left[ -\frac{\hbar ^{2}}{2M}%
\nabla ^{2}+\left( g_{11}\left\vert \psi _{+}\right\vert
^{2}+g_{12}\left\vert \psi _{-}\right\vert ^{2}\right) +V_{\mathrm{SO}}+V_{%
\mathcal{PT}}({\mathbf{r}})\right.  \notag \\
\left. +\int {\mathrm{d}^{3}{\mathbf{r}}^{\prime }\left( U_{11}\left\vert
\psi _{+}\right\vert ^{2}+U_{12}\left\vert \psi _{-}\right\vert ^{2}\right) }%
\right] \psi _{+},  \notag \\
i\hbar \frac{\partial \psi _{-}}{\partial t}=\left[ -\frac{\hbar }{2M}\nabla
^{2}+\left( g_{21}\left\vert \psi _{-}\right\vert ^{2}+g_{22}\left\vert \psi
_{+}\right\vert ^{2}\right) -V_{\mathrm{SO}}+V_{\mathcal{PT}}({\mathbf{r}}%
)\right.  \notag \\
\left. +\int {\mathrm{d}^{3}{\mathbf{r}}^{\prime }\left( U_{21}\left\vert
\psi _{-}\right\vert ^{2}+U_{22}\left\vert \psi _{+}\right\vert ^{2}\right) }%
\right] \psi _{-},  \label{equation}
\end{gather}

\noindent where $\psi =(\psi _{+},\psi _{-})$ are two components of the
spinor wave functions, ${\mathbf{r}}=(x,y,z)$, $\nabla ^{2}={\partial ^{2}}/{%
\partial x^{2}}+{\partial ^{2}}/{\partial y^{2}}+{\partial ^{2}}/{\partial
z^{2}}$, $M$ is the atomic mass, and $g_{ij}(i,j=1,2)$ are strengths of
local intra- and inter-component contact interactions. In this work, we
consider local interactions which are symmetric with respect to the
components, with $g_{11}=g_{22}=g_{12}=g_{21}=4\pi \hbar ^{2}a_{s}/M$, where
$a_{s}$ is $s$-wave scattering strength. The Rashba-type SOC operator with
strength $\lambda $ is represented by $V_{\mathrm{SO}}\psi _{+}=i\lambda
\left( \partial \psi _{-}/\partial x-i\partial \psi _{-}/\partial y+\partial
\psi _{+}/\partial z\right) $ and $V_{\mathrm{SO}}\psi _{-}=i\lambda \left(
\partial \psi _{+}/\partial x+i\partial \psi _{+}/\partial y-\partial \psi
_{-}/\partial z\right) $. The $\mathcal{PT}$-symmetric potential is taken as
\begin{equation}
V_{\mathcal{PT}}({\mathbf{r}})=p_{r}\left[ \cos ^{2}(\omega x)+\cos
^{2}(\omega y)+\cos ^{2}(\omega z)\right] +ip_{i}\left[ \sin (\omega x)+\sin
(\omega y)+\sin (\omega z)\right] ,  \label{PT potential}
\end{equation}

\noindent where $p_{r}$ and $p_{i}$ are strengths of its real and imaginary
parts, which are even and odd functions of the coordinates, respectively. As
shown below, the presence of the real part of the potential is necessary for
the effective stability (robustness) of the 3D solitons, while the imaginary
part actually produces a destabilizing effect.

The nonlocal RRI potential in the last term of Eq.~(\ref{equation}) is
written as $U_{jk}(\mathbf{r}^{\prime }-\mathbf{r})=C_{6}^{(jk)}\left/
\left( R_{b}^{6}+\left\vert \mathbf{r}^{\prime }-\mathbf{r}\right\vert
^{6}\right) \right. $, with $U_{12}=U_{21}$, where $C_{6}^{(jk)}$ are
effective coefficients of the van der Waals interaction, $R_{b}$ being the
Rydberg-blockade radius. Further, we adopt dimensionless units by means of
scaling variables with respect to the relevant length and time units, $R_{b}$
and $\tau _{b}=R_{b}^{2}M/\hbar $. Then Eq. (\ref{equation}) is replaced by
the normalized GPE,

\begin{gather}
i\frac{\partial \psi _{+}}{\partial t}=\left[ -\frac{1}{2}\nabla ^{2}+\gamma
\left( \left\vert \psi _{+}\right\vert ^{2}+\left\vert \psi _{-}\right\vert
^{2}\right) +V_{\mathrm{SO}}+V_{\mathcal{PT}}({\mathbf{r}})\right.  \notag \\
\left. +\int {\mathrm{d}^{3}{\mathbf{r}}^{\prime }\tilde{U}({\mathbf{r}}-{%
\mathbf{r}}^{\prime })\left( \alpha _{11}\left\vert \psi _{+}\right\vert
^{2}+\alpha _{12}\left\vert \psi _{-}\right\vert ^{2}\right) }\psi _{+}%
\right] ,  \notag \\
i\frac{\partial \psi _{-}}{\partial t}=\left[ -\frac{1}{2}\nabla ^{2}+\gamma
\left( \left\vert \psi _{-}\right\vert ^{2}+\left\vert \psi _{+}\right\vert
^{2}\right) -V_{\mathrm{SO}}+V_{\mathcal{PT}}({\mathbf{r}})\right.  \notag \\
\left. +\int {\mathrm{d}^{3}{\mathbf{r}}^{\prime }\tilde{U}({\mathbf{r}}-{%
\mathbf{r}}^{\prime })\left( \alpha _{21}\left\vert \psi _{-}\right\vert
^{2}+\alpha _{22}\left\vert \psi _{+}\right\vert ^{2}\right) }\right] \psi
_{-}.  \label{scaled equation}
\end{gather}%
\noindent Here $\tilde{U}({\mathbf{r}}-{\mathbf{r}}^{\prime })=1/\left[ 1+({%
\mathbf{r}}-{\mathbf{r}}^{\prime })^{6})\right] $ is the RRI kernel,
dimensionless RRI coefficients are
\begin{equation}
\alpha _{jk}=-MNC_{6}^{(jk)}/(\hbar ^{2}R_{b}^{4}),  \label{alpha}
\end{equation}%
the contact-interaction strength is set as $\gamma =1$ by scaling, along
with $M=\hbar =1$, and $N$ is the number of atoms in this system, defined as
$N=N_{+}+N_{-}$, where $N_{+}=\int {\left\vert \psi _{+}({\mathbf{r}}%
)\right\vert ^{2}\mathrm{d}^{3}{\mathbf{r}}}$ and $N_{-}=\int {\left\vert
\psi _{-}({\mathbf{r}})\right\vert ^{2}\mathrm{d}^{3}{\mathbf{r}}}$ are
norms of the two components of wave function.

The system's energy, which corresponds to Eq. (\ref{scaled equation}), is
\begin{equation}
\begin{split}
E_{\mathrm{tot}}& =\int {\ \left( \varepsilon _{k}+\varepsilon _{\mathrm{SOC}%
}+\varepsilon _{\mathrm{local}}+\varepsilon _{\mathrm{SC}}+\varepsilon _{%
\mathrm{ext}}\right) \mathrm{d}^{3}{\mathbf{r,}}} \\
\varepsilon _{k}& =\frac{1}{2}\left( \left\vert \nabla \psi _{+}\right\vert
^{2}+\left\vert \nabla \psi _{-}\right\vert ^{2}\right) , \\
\varepsilon _{\mathrm{SOC}}& =\lambda \left[ \psi _{+}^{\ast }\left( {\frac{{%
\partial \psi _{-}}}{{\partial x}}}-i{\frac{{\partial \psi _{-}}}{{\partial y%
}}}+{\frac{{\partial \psi _{+}}}{{\partial z}}}\right) -\psi _{-}^{\ast
}\left( {\frac{{\partial \psi _{+}}}{{\partial x}}}+i{\frac{{\partial \psi
_{+}}}{{\partial y}}}-{\frac{{\partial \psi _{-}}}{{\partial z}}}\right) %
\right] , \\
\varepsilon _{\mathrm{local}}& =\frac{1}{2}\gamma \left( |\psi
_{+}|^{4}+|\psi _{-}|^{4}+2|\psi _{+}|^{2}|\psi _{-}|^{2}\right) , \\
\varepsilon _{\mathrm{RRI}}& =\frac{1}{2}|\psi _{+}|^{2}\int {\mathrm{d}^{3}{%
\mathbf{r}}^{\prime }\left( \alpha _{11}\tilde{U}_{11}\left\vert \psi
_{+}\right\vert ^{2}+\alpha _{12}\tilde{U}_{12}\left\vert \psi
_{-}\right\vert ^{2}\right) } \\
& +\frac{1}{2}|\psi _{-}|^{2}\int {\mathrm{d}^{3}{\mathbf{r}}^{\prime
}\left( \alpha _{21}\tilde{U}_{21}\left\vert \psi _{-}\right\vert
^{2}+\alpha _{22}\tilde{U}_{22}\left\vert \psi _{+}\right\vert ^{2}\right) ,}
\\
\varepsilon _{\mathrm{ext}}& =\frac{1}{2}V_{\mathcal{PT}}({\mathbf{r}}%
)\left( |\psi _{+}|^{2}+|\psi _{-}|^{2}\right) ,
\end{split}
\label{energy equation}
\end{equation}%
\noindent where $\varepsilon _{k}$, $\varepsilon _{\mathrm{SOC}}$, $%
\varepsilon _{\mathrm{local}}$, $\varepsilon _{\mathrm{RRI}}$, and $%
\varepsilon _{\mathrm{ext}}$ represent the kinetic, SOC, local (contact),
nonlocal (RRI), and external potential energies, respectively. The above
condition $\gamma =1$ and Eq. (\ref{alpha}) imply that we consider the
system with local repulsion, while the long-range RRI may be both attractive
and repulsive. In the case when all nonlinear interactions are repulsive and
the periodic potential is present, multidimensional self-trapped states may
exist as gap solitons \cite{book}.

For experimental consideration, this model can be realized, in particular,
in an ultracold gas of $^{88}\mathrm{Sr}$ atoms \cite%
{Rydberg2019BaiZhengyang}. In that case, relevant values of the physical
parameters are $C_{6}=2\pi \times 81.6~\mathrm{GHz}\cdot \mathrm{\mu m}^{6}$%
, $R_{b}=10~\mathrm{\mu m}$, and $\tau =0.18~\mathrm{s}$.

\section{The results and discussions}

Stationary solution of Eq. (\ref{scaled equation}) are looked for in the
usual form, $\psi _{\pm }=\phi _{\pm }e^{-i\mu t}$, where $\mu $ is the
chemical potential, and $\phi _{\pm }$ are components of the stationary wave
function. In the polar coordinates ($r,\theta $), the initial ansatz \cite{SOC2015Stable,SOC-Zhao} for the
3D SV mode is
\begin{equation}
\phi _{\pm }=A_{\pm }r^{S_{\pm }}\exp (-a_{\pm }r^{2}+iS_{\pm }\theta ),
\label{SV}
\end{equation}%
\noindent and for the MM one it is
\begin{equation}
\phi _{\pm }=A_{1}r^{\left\vert S_{1}\right\vert }\exp (-a_{1}r^{2}\pm
iS_{1}\theta )\mp A_{2}r^{\left\vert S_{2}\right\vert }\exp (-a_{2}r^{2}\mp
iS_{2}\theta ).  \label{MM}
\end{equation}%
\noindent Here $S_{-}=S_{+}+1$ and $S_{2}=S_{1}+1$ are the corresponding
vorticities, $A_{\pm }$, $A_{1}$ and $A_{2}$ are amplitudes of the
components, and $a_{\pm }$, $a_{1}$, and $a_{2}$ are positive parameters.
Below, the components corresponding to signs $+$ and $-$ in Eqs. (\ref{SV})
and (\ref{MM}) are referred to as SV+, SV- and MM+, MM-, respectively.

As mentioned above, the fundamental SV states are produced by $(S_{+},S_{-})=(-1,0)$ and $(S_{+},S_{-})=(0,1)$, while the ESs correspond to the other integer values of $S_{+}$ and $S_{-}$. Similarly, for the MM-type ansatz, the fundamental states correspond to $(S_{1},S_{2})=(-1,0)$ and $(S_{1},S_{2})=(0,1)$. 
One component with topological charge equal to $0$ represents the fundamental solitons, and the other component represents the excited state. Thus, the whole state is called fundamental state. For the other cases, none of the component possess zero topological charge, the two components are both excited states, and the whole state is called excited states. The differences between the fundamental and excited states lie in the norm. Generally, the solitons with fundamental state has lower norm.

In the 3D SOC system, stable 3D solitons are hard to generate. Nevertheless,
3D robust (quasi-stable) solitons can be obtained in the system including
RRI and the $\mathcal{PT}$-symmetric potential. The 3D SV and MM solitons
can be generated and exist for a finite time, but they undergo nonvanishing
evolution and eventually collapse at large times. The robustness of these
solitons can be quantified by changes of their norm and shape during the
evolution. Robust 3D solitons of the SV and MM types, both fundamental (with
$S_{+,1}=0$) and ES ones (with $S_{+,1}\geq 1$), are found by means of the
imaginary-time integration method \cite{Bao,AITE2008YangJianke}. Typical
examples of the 3D robust solitons of these types are shown in Figs.~\ref%
{fig SV-soliton} and \ref{fig MM-soliton}, where the soliton profiles are
obtained from the analytical ans\"{a}tze (\ref{SV}) and (\ref{MM}).

\subsection{The robustness and evolution of the 3D SV and MM solitons}

\begin{figure}[tbph]
\centering
\includegraphics[width=0.95\textwidth]{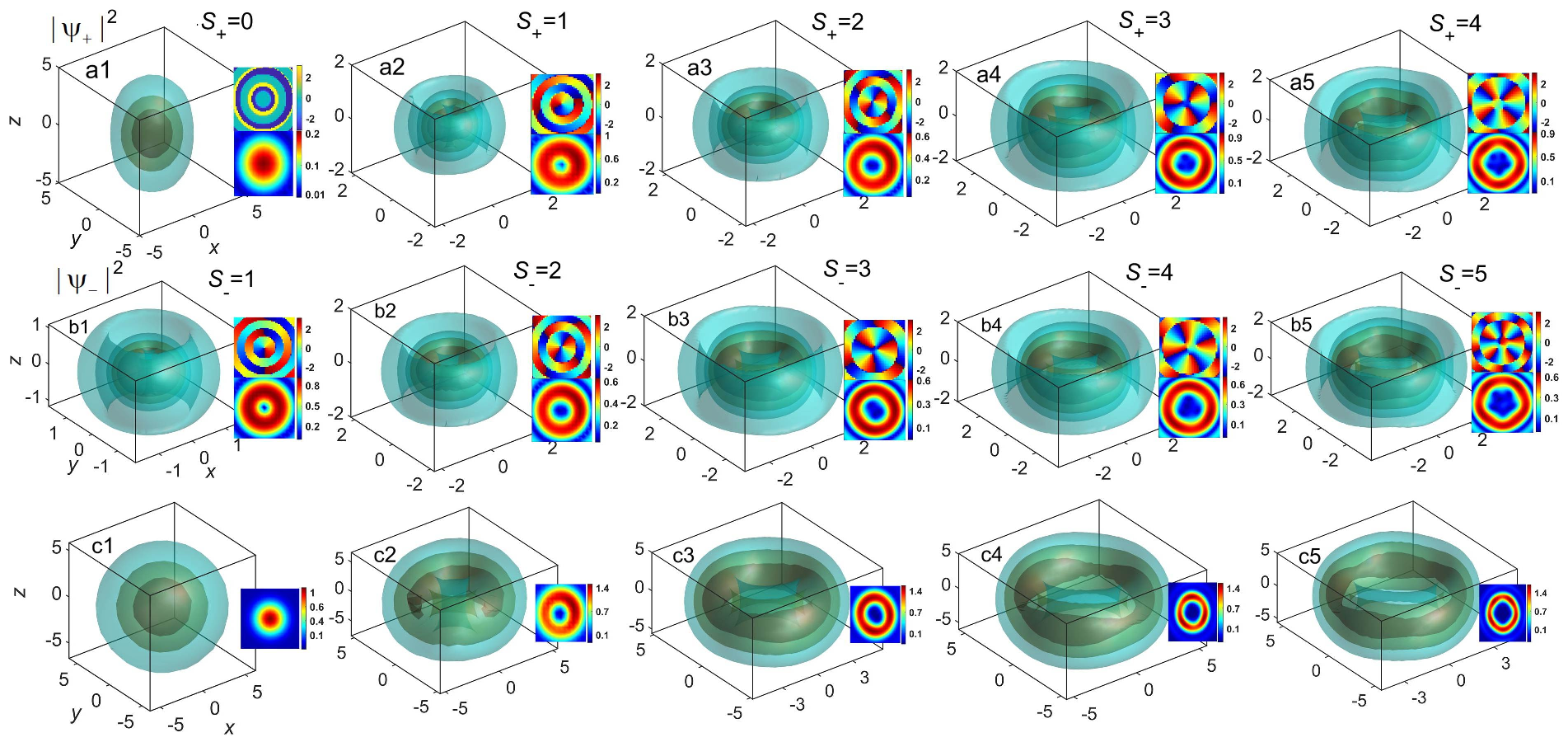}\vskip-0pc
\caption{The density distribution in the SV type solitons. The intensities
of three isosurface layers correspond to $85\%$, $50\%$, and $5\%$ of the
maximum value (in other figures density isosurfaces are plotted according to
the same definition). The first to third lines display, respectively, the
SV+ sand SV- components, and the total density of both components. The
vorticity $S_{+}$ from the first to fifth columns are 0$\sim $4, and $%
S_{-}=S_{+}+1$. The insets are the corresponding phase and density
distributions in the ($x,y$) plane. The parameters for the SVs are given in
Tab.~\protect\ref{table parameters}. }
\label{fig SV-soliton}
\end{figure}

\begin{figure}[tbph]
\centering
\includegraphics[width=0.95\textwidth]{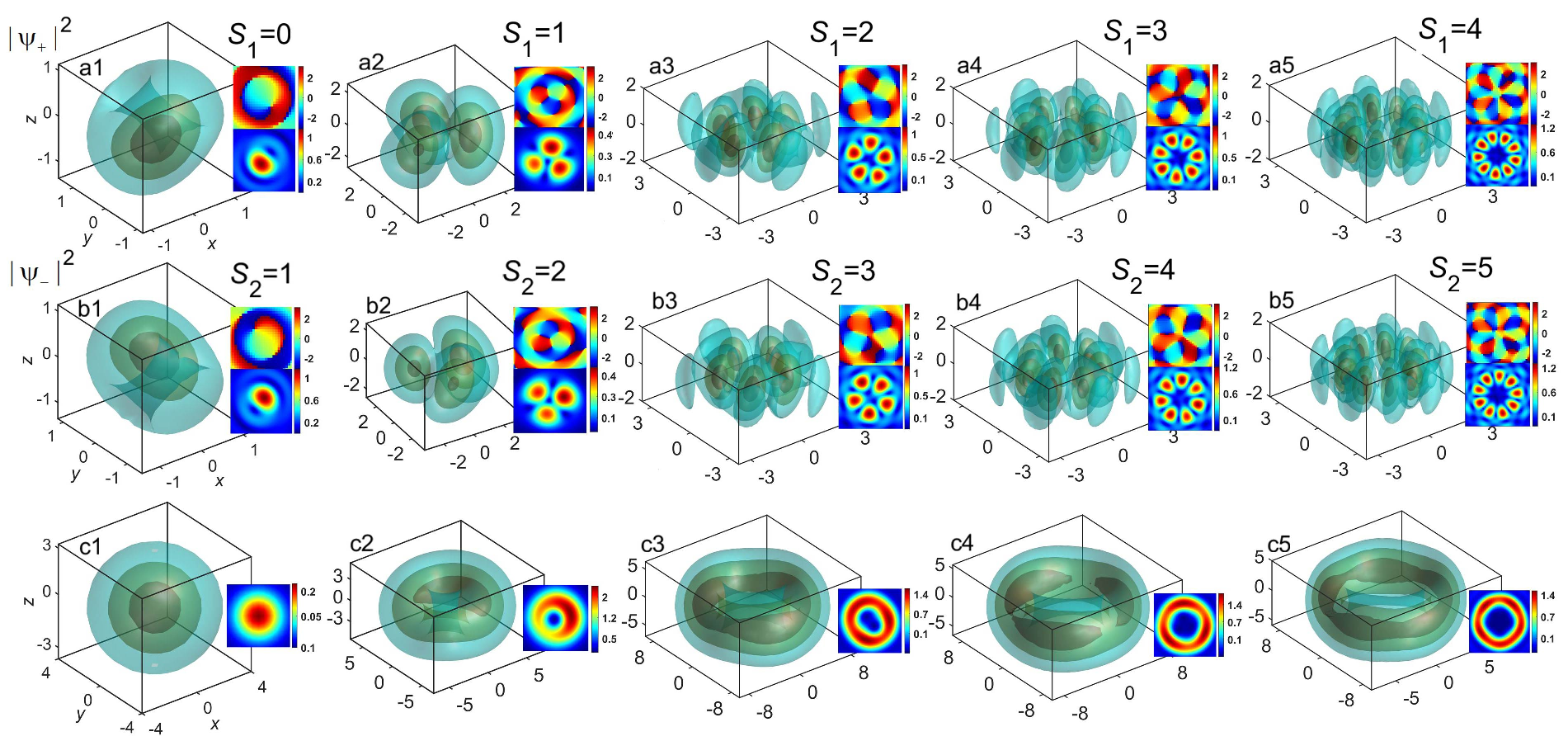}\vskip-0pc
\caption{MM solitons with MM+ and MM-types. The topological charges from the
first to the fourth column are $S_{1}=0,1,2,3,4$ and $S_{2}=S_{1}+1$. The
first to the fifth rows are MM+, MM- components, and total modes,
respectively. The parameters for MMs are given in Tab.~\protect\ref{table
parameters}. }
\label{fig MM-soliton}
\end{figure}

\begin{figure}[tbph]
\centering
\includegraphics[width=0.95\textwidth]{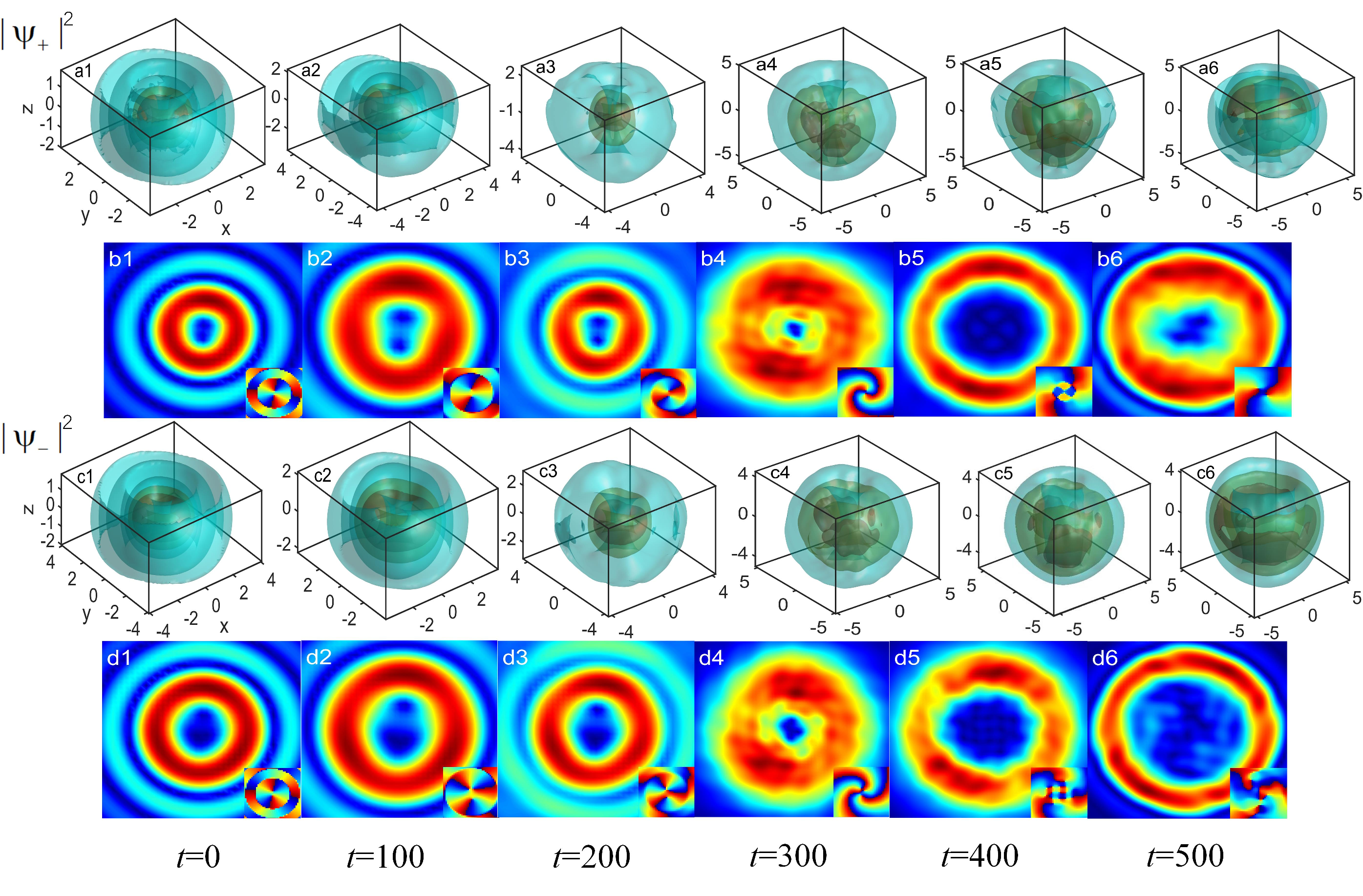}\vskip-0pc
\caption{The evolution of the density and phase distributions in two
components of a typical 3D\ SV soliton with $S_{+}=2$. The snapshots are
displayed at times $t=0,100,200,300,400,500$. The other parameters are: $%
\protect\lambda =3$, $p_{r}=-0.1$, $p_{i}=0.02$, and $\protect\alpha _{11}=%
\protect\alpha _{12}=\protect\alpha _{21}=\protect\alpha _{22}=0.1$. }
\label{fig evolution}
\end{figure}

For $(S_{+},S_{-})=(0,1)$, the SV+ and SV- components are the fundamental
and vortical ones, as shown in panels (a1) and (b1) of the first column of
Fig.~\ref{fig SV-soliton}. Vorticities $(S_{+},S_{-})$ of the ESs are $(1,2)$%
, $(2,3)$, $(3,4)$, and $(4,5)$ in the second to fifth columns in Fig.~\ref%
{fig SV-soliton}. For ESs, both components of SVs are vortical ones. The
simulations have demonstrated that SOC BEC system can support robust 3D SV
solitons with vorticities $S_{+}\leq 4$. A typical example of the soliton
evolution for $S_{+}=2$, performed by means of the split-step Fourier method
up to $t=500$, is presented in Fig. \ref{fig evolution}. In the course of
this simulation, the total norm of the SV soliton is strictly conserved. The
consideration of Fig. \ref{fig evolution} demonstrates that the SV soliton
stays fully robust up $t=300$. In physical units, this time is $\simeq 40$
s, which is definitely sufficient for the experimental observation. The
parameters for robust SVs in Fig. \ref{fig SV-soliton} are presented in Tab.~%
\ref{table parameters}, where the control parameters are SOC strength $%
\lambda $, amplitudes of the real and imaginary parts of the $\mathcal{PT}$%
-symmetric potential (\ref{PT potential}), $p_{r}$ and $p_{i}$, and RRI
strengths $\alpha _{jk}$.

\begin{figure}[tbph]
\centering
\includegraphics[width=0.95\textwidth]{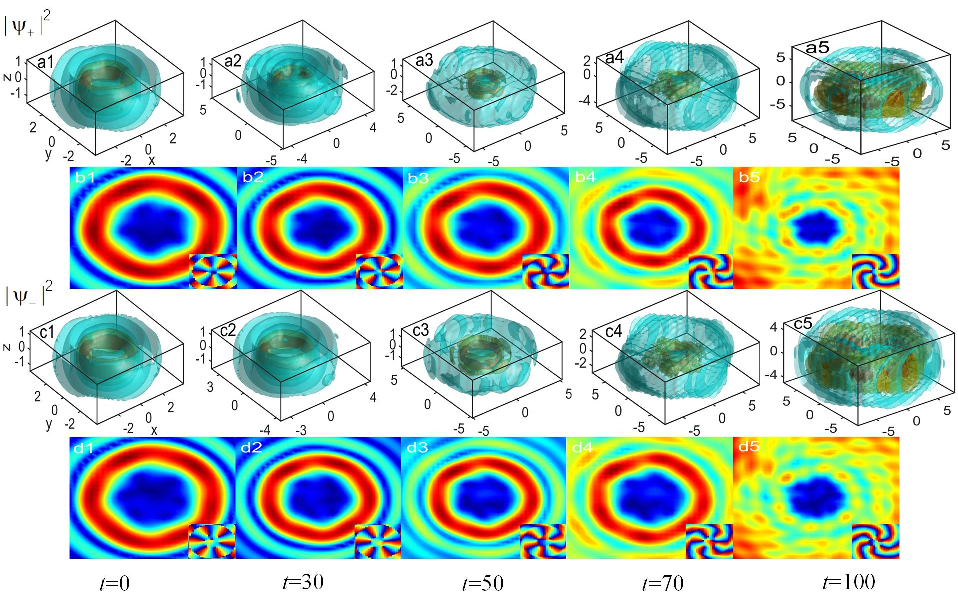}\vskip%
-0pc
\caption{The evolution of typical example of unrobust 3D\ SV solitons with $%
(S_{+}, S_{-})=(5, 6)$. The snapshots are displayed at times $%
t=0,10,50,70,100$. The other parameters are the same as Fig.~\protect\ref%
{fig evolution}.}
\label{fig evolution-unstable}
\end{figure}

At $S_{+}>4$, the SVs are not robust anymore. A typical example of the
evolution of an unstable SV, with vorticities $(S_{+},S_{-})=(5,6)$, is
shown in Fig.~\ref{fig evolution-unstable}. In this case, the SV lasts until
$z=50$ ($\simeq 9$ s, in physical units), which is significantly shorter
than in the case of the robust SV solitons with $S_{+}\leq 4$.

For the MM ansatz, the system is also capable of supporting both fundamental
and ES solitons in 3D. The respective density distributions are plotted in
Fig.~\ref{fig MM-soliton}. The fundamental state is presented in the first
column, corresponding to vorticities $(S_{1},S_{2})=(0,1)$. The subsequent
columns, ranging from the second to the fifth one, show ESs with $%
(S_{1},S_{2})=(1,2),(2,3),(3,4),(4,5)$.

In contrast to SVs, MMs exhibit quasi-discrete necklace-shaped distributions
around a central ring. The number of the discrete cores is given by $%
N_{c}=2S_{1}+1$, with $S_{1}$ denoting the vorticity of the MM+ component.
Note that the densities of the MM+ and MM- components are shaped in
alternating patterns along the ring, hence the total density distribution of
the MM solitons features a smooth ring-like configuration.

\begin{figure}[tbph]
\centering
\includegraphics[width=0.95\textwidth]{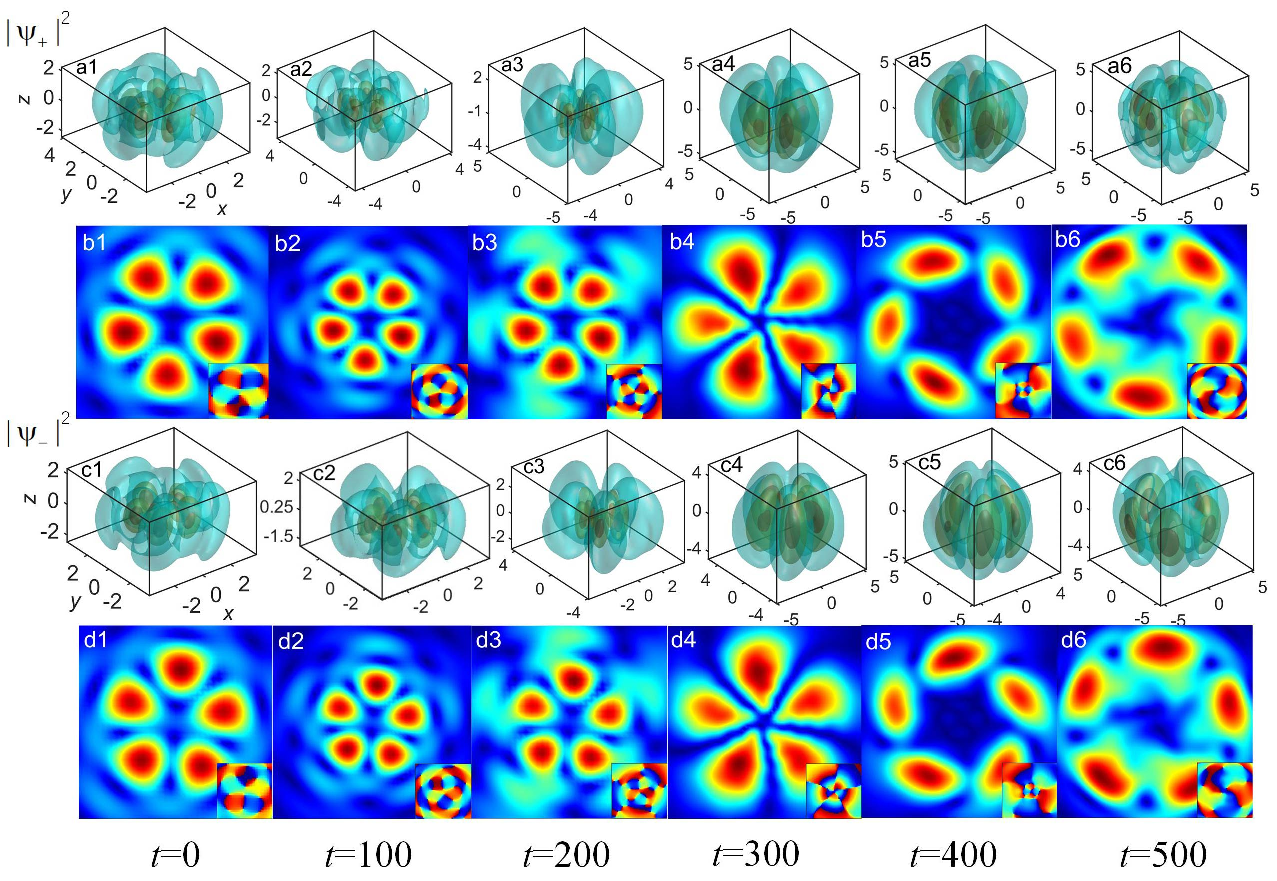}\vskip-0pc
\caption{The same as in Fig. \protect\ref{fig evolution}, but for a typical
3D soliton of the MM type with vorticity $S_{1}=2$. The parameters are the
same as in Fig. \protect\ref{fig evolution}. }
\label{fig MM-evolution}
\end{figure}

A typical example of the robust evolution of a 3D MM soliton with vorticity $%
S_{1}=2$ is displayed in Fig. \ref{fig MM-evolution}. As well as in the case
of the SV soliton shown in Fig. \ref{fig evolution}, the evolution is shown
up to $t=500$, and it can be concluded that the it stays fully robust up to $%
t=300$, i.e., time $\simeq 40$ s in physical units, Also similar to the
result for the SV soliton, the norm of the MM one is strictly conserved in
the course of the evolution.

\begin{table}[tbp]
\caption{Parameters for robust SVs and MMs of the fundamental and ES types.}
\label{table parameters}\centering%
\begin{tabular}{ccccc}
\hline
Topological charge & $\lambda $ & $p_{r}$ & $p_{i}$ & $\alpha _{jk}$ \\
\hline
SVs ($S_{+}$, $S_{-}$) &  &  &  &  \\
(0, 1) & 3.8 & -0.1 & 0.01 & 1 \\
(1, 2) & 4 & -0.1 & 0.02 & 1 \\
(2, 3) & 3 & -0.1 & 0.02 & 1 \\
(3, 4) & 3 & -0.1 & 0.01 & 1 \\
(4, 5) & 3.5 & -0.1 & 0.01 & 1 \\
MMs ($S_{1}$, $S_{2}$) &  &  &  &  \\
(0, 1) & 2.8 & -0.1 & 0.01 & 1 \\
(1, 2) & 3.2 & -0.1 & 0.01 & 1 \\
(2, 3) & 2.3 & -0.1 & 0.02 & 1 \\
(3, 4) & 2.7 & -0.1 & 0.02 & 1 \\
(4, 5) & 3.3 & -0.1 & 0.01 & 1 \\ \hline
\end{tabular}%
\end{table}

\begin{figure}[tbph]
\centering
\includegraphics[width=0.95\textwidth]{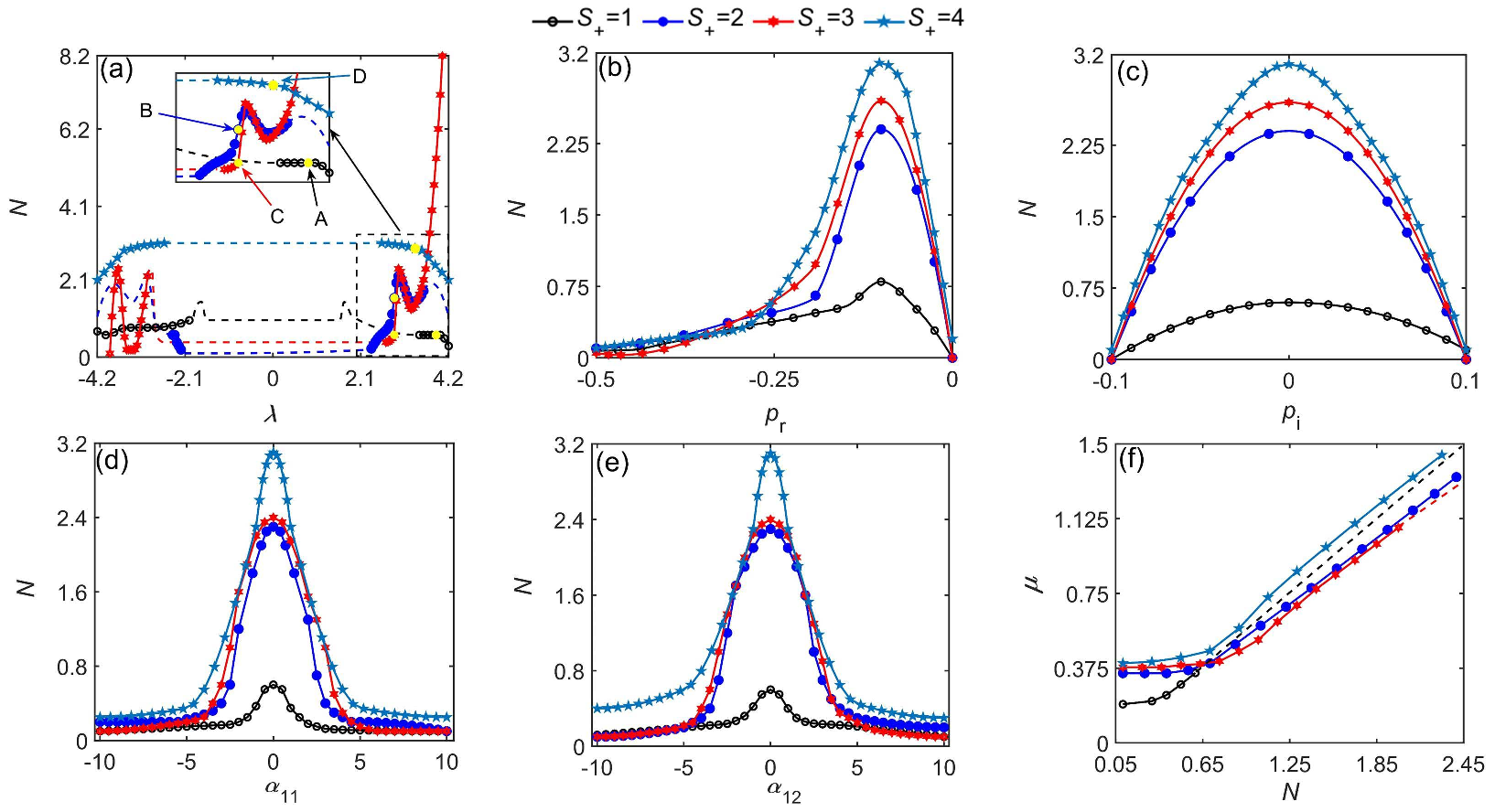}\vskip-0pc
\caption{The norm of the SV-type solitons as a function of (a) SOC strength $%
\protect\lambda $, amplitudes pf real (b) and imaginary (c) parts of the $%
\mathcal{PT}$-symmetric potential (\protect\ref{PT potential}), (d) intra-
and (e) inter-component RRI strength, (f) chemical potential $\protect\mu $.
The inset in (a) is the zoom of region $\protect\lambda \in (2.0~4.2)$,
where points A, B, C and D designate states correspond to the SVs with
vorticities $S_{+}=1,2,3,4$ in Fig. ~\protect\ref{fig SV-soliton}. Solid and
dashed lines represent robust and strongly unstable SVs, respectively. The
fixed parameters for each panel are presented in Tab. \protect\ref{table
parameters}. }
\label{fig SV-line}
\end{figure}

The modulation of SVs following the variation of the system's parameters,
such as the SOC constant $\lambda $, amplitudes of the real and imaginary
parts of the potential, intra- and inter-component of RRI strengths, as well
as the dependence between the chemical potential $\mu $ and total norm, are
shown in Fig.~\ref{fig SV-line}. The $N(\lambda )$ relations displayed in
Fig.~\ref{fig SV-line}(a) delineate robustness domains for the SV solitons
for different vorticities. It is seen that the domains are divided in two
distinct regions, situated, respectively, at $\lambda <0$ and $\lambda >0$,
indicating this system cannot support robust SVs unless the SOC strength $%
|\lambda |$ exceeds a certain threshold value. Details of the robustness
domains are summarized in Tab.~\ref{table stable domains}. SVs do not remain
either if the SOC strength is too large, \textit{viz}., at $\left\vert
\lambda \right\vert >4.2$.

For different vorticities, the SV norms exhibit different behavior for the
variation of $\lambda $. For $S_{+}=1$ and $4$, the norm changes smoothly
with $\lambda $, while it strongly fluctuates for $S_{+}=2$ and $3$. For
example, $N$ changes by $\simeq 10$ times when $\lambda $ varies from $3.2$
to $4.2$ for $S_{+}=3$. The increase of the vorticity enhances the
fluctuations of the norm, eventually leading to full instability of SVs at $%
S_{+}\geq 5$ (as mentioned above). The parameter dependence for the 3D MM
solitons is similar to that for SVs (not shown here in detail).

The dependence of the gap of values of the SOC constant$\ \lambda $, which
divides the robustness domain at $\lambda <0$ and $\lambda >0$, on
parameters $p_{i}$ and $\alpha _{12}$ is shown in Fig. \ref{fig combination
domain}. Here $S_{+}=3$ for SVs and $S_{1}=3$ for MMs, the other parameters
being $p_{r}=-0.1$, $\alpha _{11}=\alpha _{22}=1$.

Amplitudes of the real and imaginary parts of the $\mathcal{PT}$-symmetric
potential are essential parameters which control the robust SV solitons.
Figures~\ref{fig SV-line}(b) and (c) display the soliton's norm vs. $p_{r}$
and $p_{i}$. It is seen that only $p_{r}<0$ can support robust SVs, as in
the opposite case the soliton is placed at a local potential maximum,
instead of the minimum (see Eq. (\ref{PT potential})). On the other hand,
there is natural symmetry between positive and negative values of $p_{i}$.
Taking Fig.~\ref{fig SV-soliton} and Tab.~\ref{table stable domains} for
example, with fixed parameters $(S_{+},\lambda ,\alpha
_{ij},p_{r})=(2,3,1,-0.1)$, one obtains the SV\ soliton solutions with real
chemical potential $\mu $ in the interval of $-0.1<p_{i}<+0.1$. In this
case, $p_{i}=\pm 0.1$ are $\mathcal{PT}$-symmetry-breaking points \cite%
{symm-breaking,parity-time2-2016} beyond which the stationary solutions are
unphysical, featuring formally complex values of $\mu $. The norm attains
its maximum for the real potential, with $p_{i}=0$, and decreases with the
increase of $\left\vert p_{i}\right\vert $. Actually, $\mathcal{PT}$
symmetry remains unbroken at $\left\vert p_{i}/p_{r}\right\vert <1$.
\begin{table}[tbp]
\caption{Robustness domains for the 3D\ SV solitons}
\label{table stable domains}\centering%
\begin{tabular}{ccccc}
\hline
$(S_{+},S_{-})$ & $\lambda $ & $p_{r}$ & $p_{i}$ & $\alpha _{jk}$ \\ \hline
(1, 2) & -4.2$\sim $ -2, 3.5$\sim $4.2 & -0.5$\sim $0 & -0.1 $\sim $ 0.1 &
-10 $\sim $ 10 \\
(2, 3) & -2.4$\sim $ -2.2, 2.3$\sim $3.6 & -0.5$\sim $0 & -0.1 $\sim $ 0.1 &
-10 $\sim $ 10 \\
(3, 4) & -3.9$\sim $ -3, 2.7$\sim $4.2 & -0.5$\sim $0 & -0.1 $\sim $ 0.1 &
-10 $\sim $ 10 \\
(4, 5) & -4.2$\sim $ -2.6, 2.6$\sim $4.2 & -0.5$\sim $0 & -0.1 $\sim $ 0.1 &
-10 $\sim $ 10 \\ \hline
\end{tabular}%
\end{table}

The norm of the SV solitons are shown, as a functions of the intra- and
inter-component RRI strengths, $\alpha _{11}$ and $\alpha _{12}$, in Figs.~%
\ref{fig SV-line}(d) and (e), respectively. It is seen that the attractive ($%
\alpha _{jk}<0$) and repulsive ($\alpha _{jk}>0$) interactions have
approximately the same effect on the modulation of the SV solitons.
Furthermore, the effect is nearly the same for the intra- and
inter-component RRI strength, $\alpha _{11}$ and $\alpha _{12}$. Note that,
quite naturally, the norm of the SV solitons nearly vanishes for very strong
RRI, $\left\vert \alpha _{ij}\right\vert >10$.

The $\mu (N)$ curves which are displayed, for different values of $S_{+}$,
in Fig.~\ref{fig SV-line}(f), satisfy the \textit{anti-Vakhitov-Kolokolov
criterion}, i.e., $\mathrm{d}\mu /\mathrm{d}N>0$, which is a known necessary
condition for the stability of bright solitons supported by self-repulsive
nonlinearities \cite{VK2010Boris} (the Vakhitov-Kolokolov criterion proper, $%
\mathrm{d}\mu /\mathrm{d}N<0$ \cite%
{VK1973Stationary,wavecollapse1-1998Wave,wavecollapse2-2011kuznetsov}, is
necessary for stability of solitons in the case of the attractive
nonlinearity).

\begin{figure}[tbph]
\centering
\includegraphics[width=0.95\textwidth]{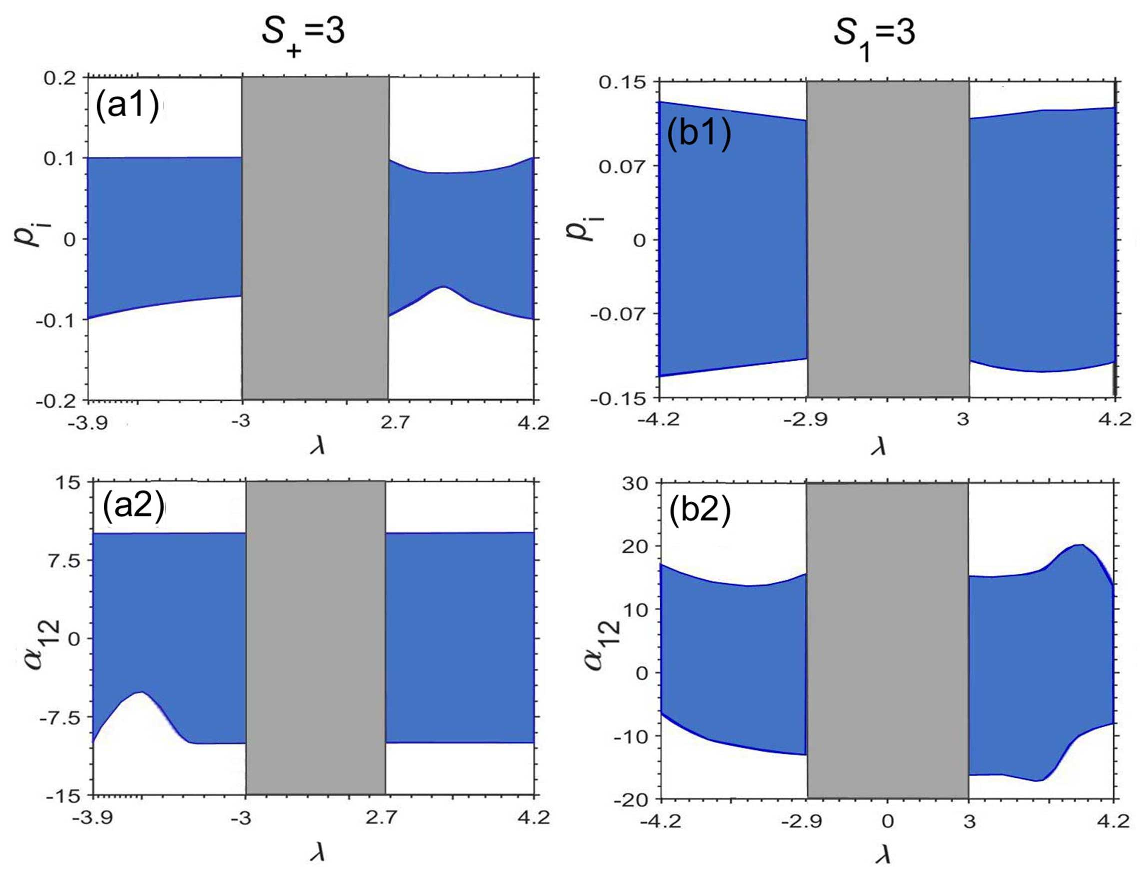}\vskip%
-0pc
\caption{Blue areas: the robustness domains in paraneter planes($\protect%
\lambda ,p_{i}$) and ($\protect\lambda ,\protect\alpha _{12}$). The first
column is for the 3D SV solitons with $S_{+}=3$, and the second column is
for the MMs with $S_{1}=3$. Gray areas represent gaps which separate the
robustness domains and $\protect\lambda <0$ and $\protect\lambda >0$. }
\label{fig combination domain}
\end{figure}

\subsection{The energy and orbital angular momentum of SVs and MMs}

\begin{figure}[tbph]
\centering
\includegraphics[width=0.95\textwidth]{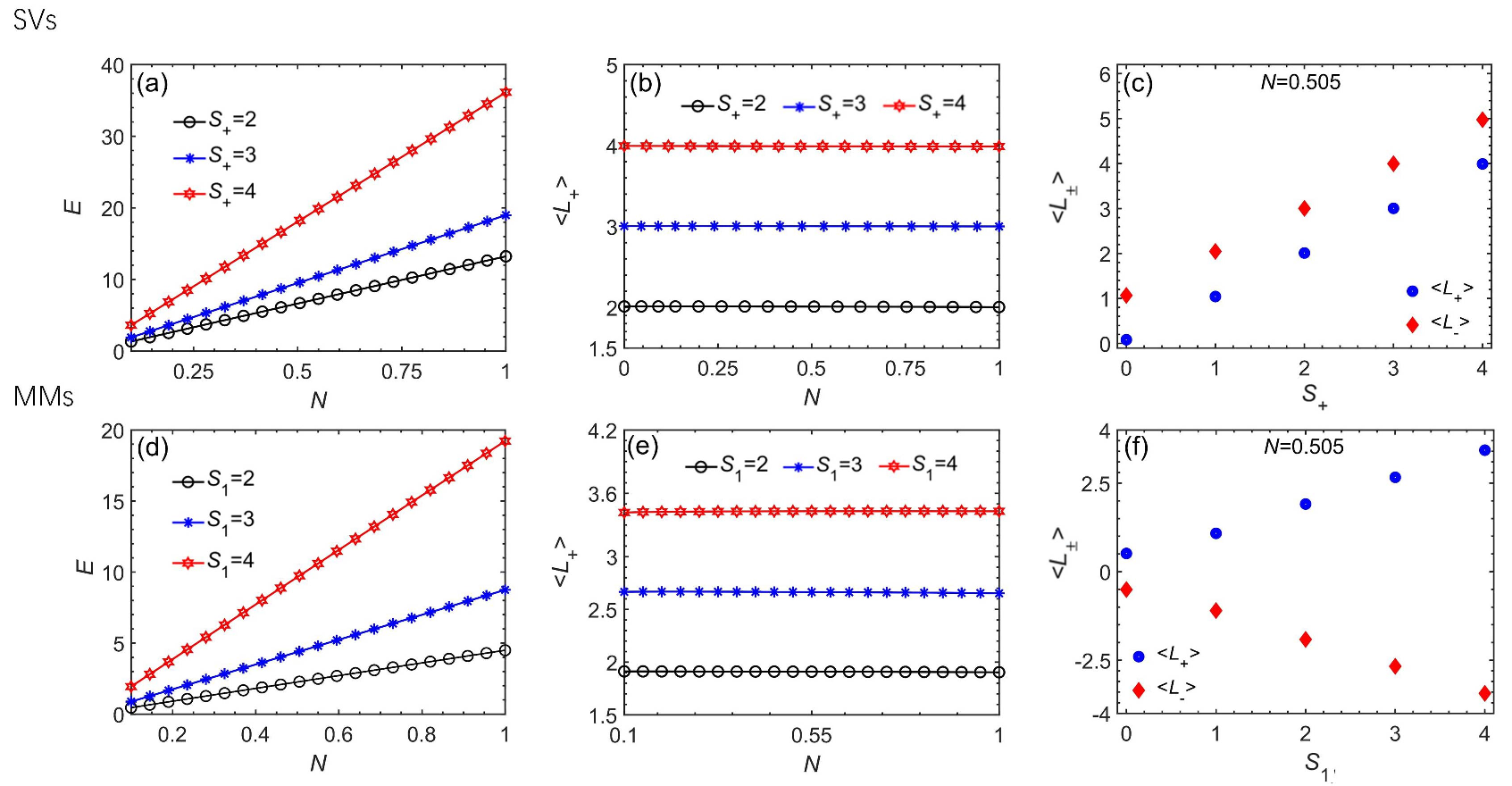}\vskip-0pc
\caption{The energy (a,d) and OAM (b,e) as functions of the total norm for
the 3D solitons of the SV and MM types. (c,f) Values of OAM for different
forticities with a fixed norm. The black, blue and red lines in panels
(a,b,d,e) correspond to vorticities $S_{+}=S_{1}=2,3,4$, respectively. The
parameters are given in Tab.~\protect\ref{table parameters}. }
\label{fig energy}
\end{figure}

The families of the 3D solitons of the SV and MM types with different
vorticities can be naturally characterized by dependences of the total
energy (\ref{energy equation}) on the total norm $N$, which are plotted in
Figs.~\ref{fig energy}(a) and (d), respectively. It is seen that the energy
increases as a function of $N$ and vorticity. Another characteristic of the
spinor solitons is the orbital angular momentum (OAM) ,

\begin{equation}
\begin{split}
\left\langle L_{\pm }\right\rangle & =\int {\mathrm{d}^{3}{\mathbf{r}}\frac{%
\psi _{\pm }^{\ast }{\hat{L}}\psi _{\pm }}{N_{\pm }}} \\
\left\langle L\right\rangle & =\frac{N_{+}\left\langle L_{+}\right\rangle
+N_{-}\left\langle L_{-}\right\rangle }{N}
\end{split}
\label{angular momentum}
\end{equation}%
\noindent where components $L_{\pm }$ should be calculated with the OAM
operator ${\hat{L}}=-i(x\partial /\partial y-y\partial /\partial x)$ \cite%
{SOC2-2018Liyongyao}. 
Figure~\ref{fig energy}(b) displays the $\left\langle L_{+}\right\rangle (N)$ relations for SVs. The orbital angular moment of atoms is proportional to the vorticity (topological charge).
Due to its definition (\ref%
{angular momentum}), $\left\langle L_{+}\right\rangle $ takes the integer
value which is equal to the vorticity, $\left\langle L_{+}\right\rangle
=S_{+}$, and it does not depend on the norm. Similarly, one obtains $%
\left\langle L_{-}\right\rangle =S_{-}$ for the second SV component, as
shown in Fig.~\ref{fig energy}(c). According to Eq.~\ref{angular momentum},
the total OAM is $\left\langle L\right\rangle =\left\langle
L_{+}\right\rangle +0.5$. Here, the energies and OAM are calculated from ans%
\"{a}tze (\ref{SV}) and (\ref{MM}).

For MMs, the energy increases as a function of the norm and vorticity, as
shown in Fig.~\ref{fig energy}(d). The two OAM components of OAM are also
independent of the norm, but are not equal to integer values of the
vorticity. Namely, the OAM of the $\psi _{+}$ component in Fig.~\ref{fig
energy}(e) is $\left\langle L_{+}\right\rangle =1.9,2.7,3.4$ for $%
S_{+}=2,3,4 $ respectively, being slightly smaller than the corresponding
vorticity $S_{1}$. The OAM of the $\psi _{-}$ component is $\left\langle
L_{-}\right\rangle =-\left\langle L_{+}\right\rangle $, hence the total MM's
OAM is zero: $\left\langle L\right\rangle =(\left\langle L_{+}\right\rangle
+\left\langle L_{-}\right\rangle )/2=0$.

\section{Conclusion}

In this work, we have introduced the 3D SOC (spin-orbit-coupled) BEC system
with long-range RRI (Rydberg-Rydberg interactions) and a spatially periodic
potential, real of $\mathcal{PT}$-symmetric one. The objective is to
construct robust solitons based on the ansatz of the SV (semi-vortex) and MM
(mixed-mode) types. The \textit{ans\"{a}tze} include vorticities which
correspond to the 3D\ SV and MM solitons in the fundamental or excited
states. By means of systematic simulations of the GPE (Gross-Pitaevskii
equation) for the mean-field spinor wave functions, it is found that these
states stay robust up to intrinsic vorticity $S=4$. These include values $%
S=2,3,4$ which correspond to ESs (excited states) of the 3D solitons of the
SV and MM types, while it is known that all ESs are unstable in the local
SOC system without RRI. The 3D solitons of the SV and MM types exhibit
toroidal and necklace-shaped density patterns, respectively. The OAM
(orbital angular momentum) of the solitons is determined by vorticities of
their components. The robust solitons keep their 3D density and phase
profiles for a long evolution time, which may exceed times available to the
experiment. However, very long evolution leads to eventual destabilization
of the solitons. The work has identified the dependence of the shape and
robustness of the SV and MM solitons, both fundamental ones and ESs, on
control parameters of the systems, such as the amplitude of the imaginary
component of the $\mathcal{PT}$-symmetric potential and strengths of the SOC
and RRI\ terms.

\hspace*{\fill}\newline
\textbf{Conflict of interest} \vspace{0.3cm}

The authors declare that they have no known competing financial interests or
personal relationships that could influence the work reported in this paper.

\hspace*{\fill}\newline
\textbf{Data availability} \vspace{0.3cm}

Numerical data generated for the research described in the article is
available on a reasonable request.

\hspace*{\fill}\newline
\textbf{Acknowledgements} \vspace{0.3cm}

This work was supported by the National Natural Science Foundation of China
(62275075), Natural Science Foundation of Hubei Province (2021CFB418,
2023AFC042), and Training Program of Innovation and
Entrepreneurship for Undergraduates of Hubei Province (S202210927003). The
work of B.A.M. was supported, in part, by the Israel Science Foundation
through grant No. 1695/22.

\end{document}